\documentclass[10pt,aip,apl,superscriptaddress,onecolumn,floatfix,preprint]{revtex4-1}         
\usepackage{graphicx}
\usepackage{dcolumn}
\usepackage{bm}
\usepackage{amsmath}
\usepackage{subfigure}
\usepackage{url}

\begin{document}

\title{Shape anisotropy of magnetic nanoparticles in (Co$_{86}$Nb$_{12}$Ta$_2$)$_x$(SiO$_2$)$_{1-x}$ composite films revealed by grazing-incidence small-angle X-ray scattering}

\author{V.\,Ukleev}
\affiliation{B. P. Konstantinov Petersburg Nuclear Physics Institute, National Research Center "Kurchatov Institute", 188300 Gatchina, Russia; \\ RIKEN Center for Emergent Matter Science (CEMS), 351-0198 Wako, Saitama, Japan}
\email{victor.ukleev@riken.jp}


\begin{abstract}
Structure of amorphous nangranular composite  (Co$_{86}$Nb$_{12}$Ta$_2$)$_x$(SiO$_2$)$_{1-x}$ films was studied by means of grazing-incidence small-angle X-ray scattering as a function of atomic concentration $x$ of the metal fraction in vicinity of the percolation threshold. It has been established that the average size of magnetic nanoparticles increases with $x$ and this increment exhibit the anisotropic character. It was found that in a whole concentration range the particles are elongated along the growth direction perpendicular to the film plane. For the lowest concentration $x=0.193$ clusters with isotropic shape and average radius of 1 nm were observed, while for higher concentrations up to $x=0.665$ the average cluster length along the film normal is $1.5-7$ nm and the cross-section is $1-2$ nm.

\end{abstract}
\maketitle

\section{Introduction}
Magnetic nanogranular composites (or nanocomposites) are promising materials that combine characteristic structural and magnetic properties of nanoparticles and thin films providing an opportunity to optimize an anisotropy and dimensionality of magnetic properties by changing the composition \cite{knobel2008superparamagnetism,yao2014magnetic}. Nanocomposites consists of metallic nanoparticles incorporated in an insulating matrix. The interest in such materials is maintained due to a wide range of controlled conductive, magnetic, optical and other physical properties, which can be used in microwave devices, spintronics, medicine and biology \cite{gerber1997magnetoresistance,Stuart,kodama1999magnetic, fujimori2006spintronics, vzutic2004spintronics,atwater2010plasmonics,suber2011approaches,lambert2014all}. High-frequency devices is one of the key applications of magnetic nanocomposites due to their exceptional soft-magnetic properties. One of the most important parameters that determines the soft-magnetic characteristics of composite is the magnetic anisotropy of nanoparticles. In the case of nanocrystalline grains the anisotropy constant is determined mostly by the crystal anisotropy of the material, while a random orientation of crystallographic axes of each particle causes a high dispersion of magnetic anisotropy axes. In case of amorphous nanoparticles as a metal phase it is possible to establish a control of both magnitude and dispersion of magnetic anisotropy fields by tuning the particles concentration in a composite material.
Previously the presence of the magnetic anisotropy perpendicular to the film plane was observed in (Co$_{86}$Nb$_{12}$Ta$_2$)$_x$(SiO$_2$)$_{1-x}$ film, and elongation of the (Co$_{86}$Nb$_{12}$Ta$_2$) magnetic nanoclusters in the film growth direction was proposed. In this work, we further investigate this feature for (Co$_{86}$Nb$_{12}$Ta$_2$)$_x$(SiO$_2$)$_{1-x}$ nanocomposite films with different concentrations of metal $x$ by means of grazing-incidence small-angle X-ray scattering (GISAXS).

\section{Experiment and discussion}

\begin{figure*}
\centering\includegraphics[width=17cm]{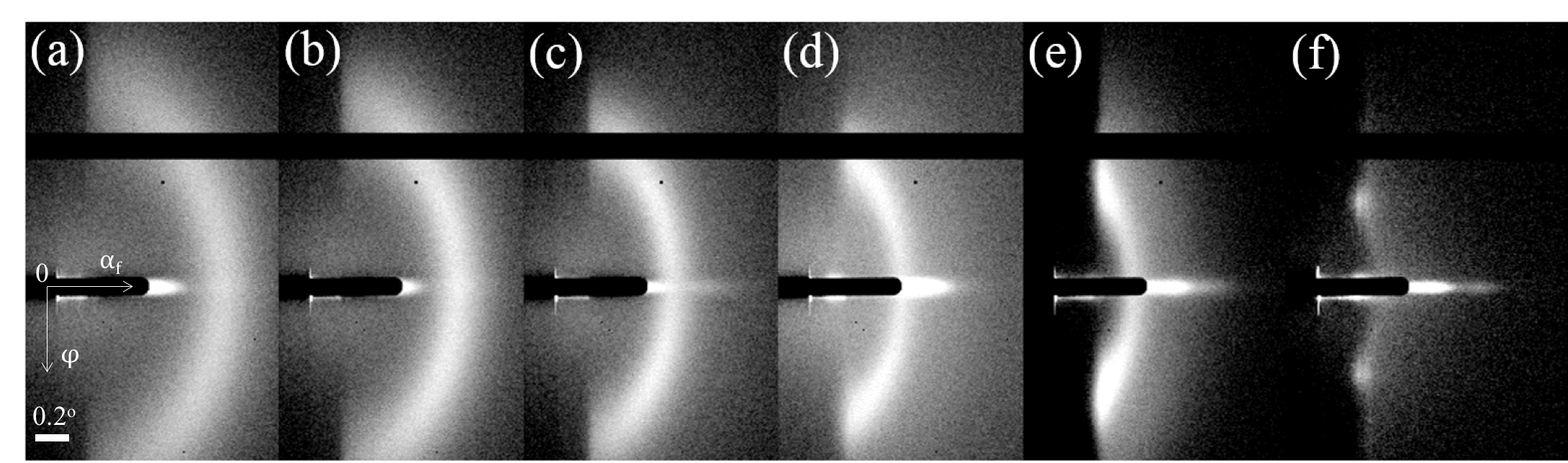}
\caption{Two-dimensional maps of GISAXS from samples (Co$_{86}$Nb$_{12}$Ta$_2$)$_x$(SiO$_2$)$_{1-x}$ with concentrations (a) $x = 0.193$, (b) 0.275, (c) 0.364 (d) 0.443, (e) 0.541, (f) 0.665.}
\label{Fig1}
\end{figure*}

\begin{figure*}[ht]
\centering\includegraphics[width=17cm]{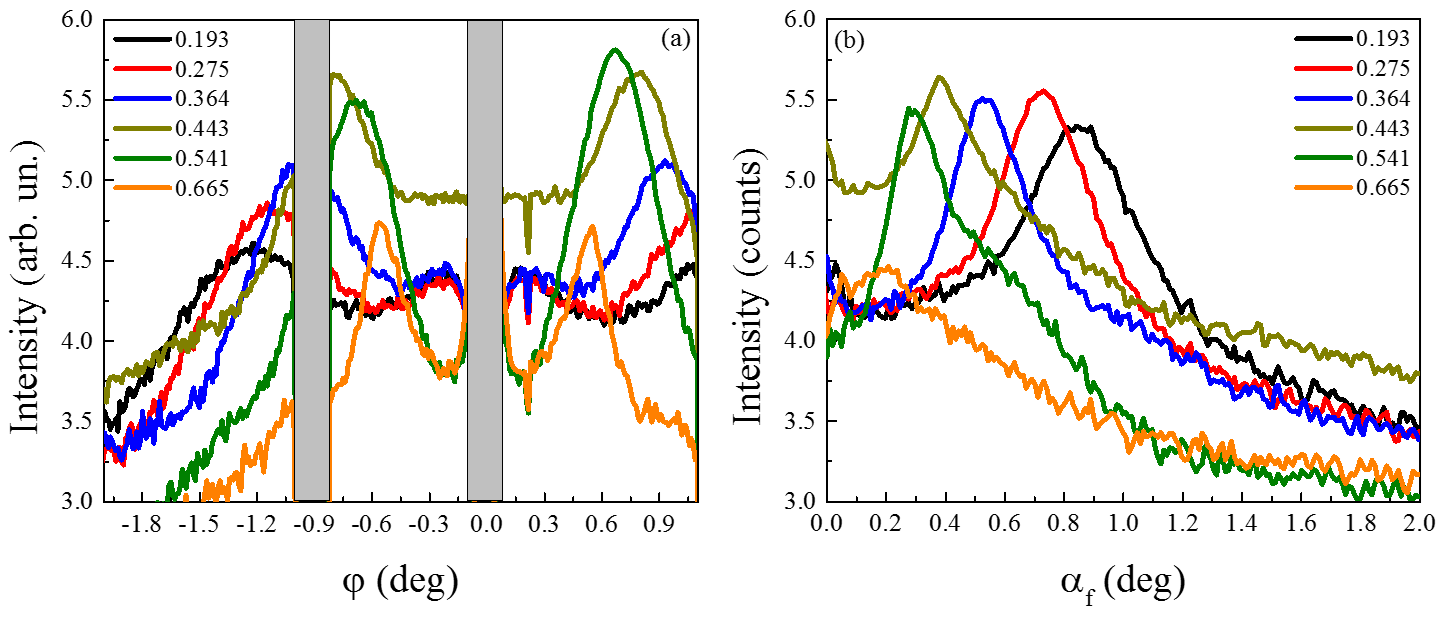}
\caption{The GISAXS intensity profiles from the samples (Co$_{86}$Nb$_{12}$Ta$_2$)$_x$(SiO$_2$)$_{1-x}$ (a) in the direction $\varphi$ for a constant value of $\alpha_f=0.2^\circ$ and (b) in the direction of $\alpha_f$ at a constant value of $\varphi=0.25^\circ$.}
\label{Fig2}
\end{figure*}

The granular nanocomposite films (Co$_{86}$Nb$_{12}$Ta$_2$)$_x$(SiO$_2$)$_{1-x}$ with a thickness of 5 $\mu$m were prepared in Voronezh State Technical University (Russia) by ion beam co-sputtering of the composite target, which consists from plates of SiO$_2$ fixed on a surface of amorphous ferromagnetic alloy Co$_{86}$Nb$_{12}$Ta$_2$. The ratio $x$ was controlled by choosing the ratio between surface areas of SiO$_2$ and metal targets. Magnetic percolation threshold in  (Co$_{86}$Nb$_{12}$Ta$_2$)$_x$(SiO$_2$)$_{1-x}$ composites corresponds to $x_p=0.46-0.48$. All peculiarities of the deposition procedure and choice of components are described in works \cite{pisarev2003magnetooptical,stognei2010anisotropy,gridnev2012nonlinear}.

Grazing-incident geometry improves the surface sensitivity compared to the transmissional small-angle X-ray scattering thus adapting this technique for investigation of surfaces and buried interfaces, thin films and multilayers. Furthermore, GISAXS is a non-destructive technique and unlike the real-space cross-sectional visualization techniques, like transmission electron microscopy, does not require any specific sample preparation. The GISAXS experiment was carried out at ID10 beamline of European Synchrotron Radiation Facility (ESRF, France). A detailed description of the GISAXS technique can be found in the review \cite{renaud2009probing}. The geometry of experiment is determined by grazing-incidence angle $\alpha_i$ of the X-ray beam on the sample surface and two scattering angles, $\alpha_f$ and $\varphi$. The angles $\alpha_i$, $\alpha_f$ and $\varphi$ determine the values of components of the momentum transfer vector: $Q_x$, $Q_y$, $Q_z$:

\begin{flalign*}
&Q_{\rm z}(\varphi , \alpha_{\rm f}) =  \frac{2\pi}{\lambda} \left( \sin \alpha_{\rm f}  + \sin \alpha_{\rm i} \right), \\
&Q_{||}(\varphi , \alpha_{\rm f}) =  \sqrt{Q_{\rm x}^2(\varphi , \alpha_{\rm f})+Q_{\rm y}^2(\varphi , \alpha_{\rm f})}, \\
&Q_{\rm x}(\varphi , \alpha_{\rm f}) =  \frac{2\pi}{\lambda} \left( \cos \alpha_{\rm f}  \cos\varphi - \cos \alpha_{\rm i} \right), \\
&Q_{\rm y}(\varphi , \alpha_{\rm f}) =  \frac{2\pi}{\lambda} \cos \alpha_{\rm f}  \sin \varphi.
\end{flalign*}

Thus, by measuring the component of the momentum transfer $Q_z$ perpendicular to the sample plane, one can get the information on the distribution of electron density in $z$ direction, while by measuring the component $ Q_{||}$ in the sample plane $(x,y)$, it is possible to study the lateral structure of the sample.

In present experiment the collimated X-ray beam with the size $10 \times 200$ $\mu$m$^2$ and the wavelength $\lambda=0.563$ \AA~falling on the surface of the sample at the grazing angle $\alpha_i= 0.25^\circ$ was used for the GISAXS experiment, what allowed to probe the sample volume depth in approximately $\sim 200$ nm below the surface. The scattering intensity was measured by the two-dimensional position-sensitive detector PILATUS-300K. The detector was protected from the hight-intensity direct and specularly reflected beams by lead absorber.

The two-dimensional GISAXS intensity maps from the sample (Co$_{86}$Nb$_{12}$Ta$_2$)$_x$(SiO$_2$)$_{1-x}$ with $x$ varying from 0.193 to 0.665 are shown in Fig. \ref{Fig1}. For all the measured samples the GISAXS patterns are typical to the disordered three-dimensional net of nanoparticles. The radius of the characteristic arc in the horizontal and vertical directions corresponds to the maxima of the correlation function in the plane of the film and perpendicular to the plane of the sample, respectively. As the concentration increases (Figures \ref{Fig1}a--f), the maximum of the scattering intensity along $\alpha_f$ shifts towards to the transmitted beam ($\alpha_f=0^\circ$), which corresponds to an increase in the size of the scattering objects in the real-space. The diffraction ring shown in Fig. \ref{Fig1}a for the sample with lowest concentration of metal $x=0.193$ corresponds to the scattering from a system of grains (magnetic nanoparticles) with the in-plane radius $r_l = 2\pi / Q_l$, where $Q_l$ is the radius of the ring along $\varphi$ direction expressed in momentum transfer units and radius perpendicular to the sample plane $r_p = 2\pi / Q_p$, where $Q_p$ is the radius of the ring along $\alpha_f$ direction, correspondingly. The scattering is almost isotropic in $(\varphi,\alpha_f)$ plane indicating isotropic shape of nanoparticles with $r_l\approx r_p$. To calculate the mean sizes of the granules for all concentrations $x$ the cross-sections of the two-dimensional intensity maps were taken along $\varphi$ direction at $\alpha_f = 0.2^\circ$ (Fig. \ref{Fig2}a) and along $\alpha_f$ direction at $\varphi=0.25^\circ$ (Fig. \ref{Fig2}b). 

\begin{figure}[ht]
\centering\includegraphics[width=8.5cm]{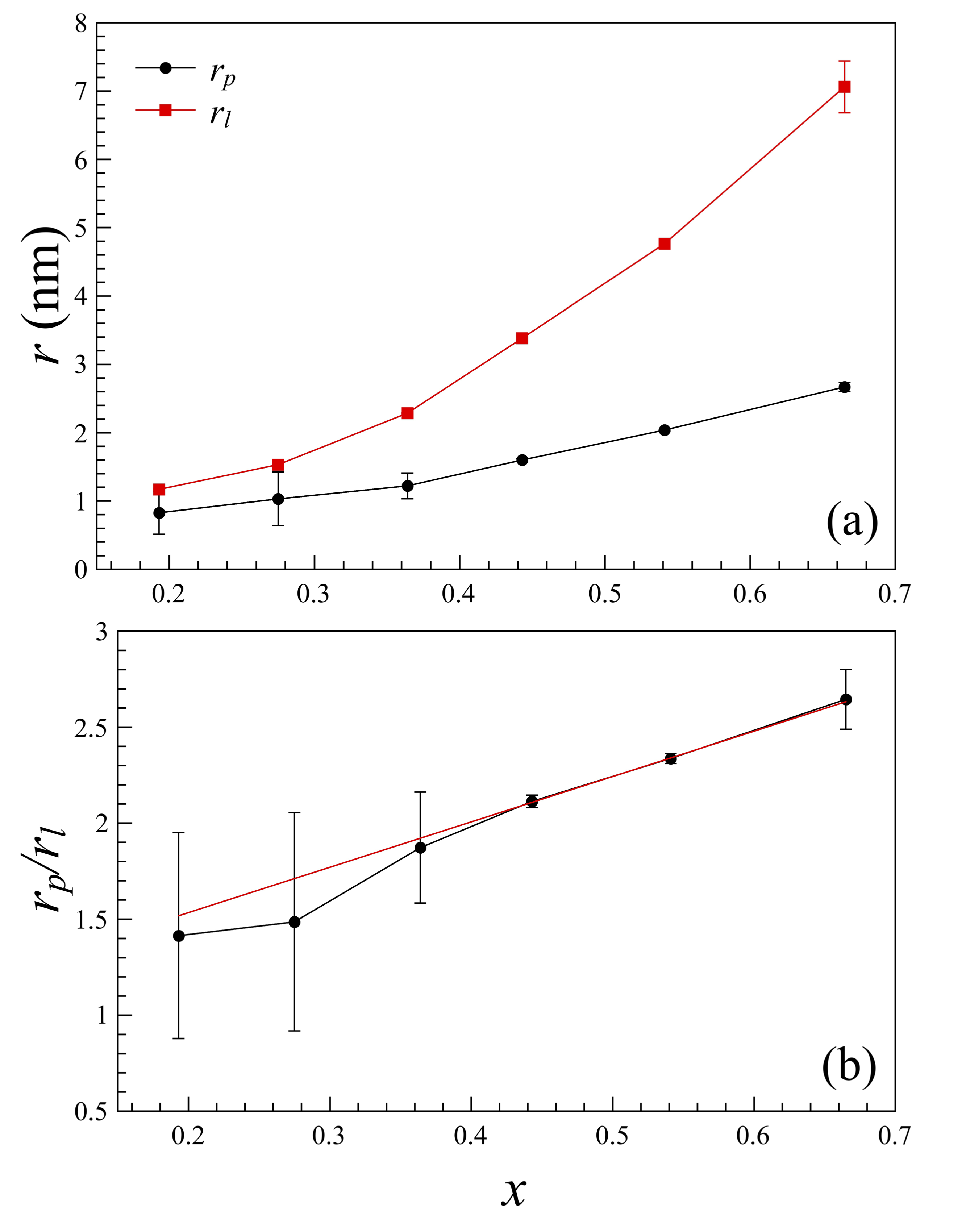}
\caption{(a) Dependence of the radius of Co$_{86}$Nb$_{12}$Ta$_2$ nanoparticles in the ($r_l$) plane and perpendicular to the film plane ($r_p$) for different values of the metal concentration and their ratio (b) that characterize shape anisotropy of the particles.}
\label{Fig3}
\end{figure}

As it seen from Fig. \ref{Fig1} the ring-shaped SAXS intensity tends to shrink in the $Q_z$ direction as the concentration $x$ increases. This is a clear indication of the Co$_{86}$Nb$_{12}$Ta$_2$ particles elongation along the $z$-axis (growth direction), while the lateral size increases moderately. The size of the nanoparticles in the in-plane of the sample $r_l$ and in the growth direction $r_p$ as a function of metal fraction $x$ are summarized in Fig. \ref{Fig3}a. Indeed, as it was judged in work \cite{stognei2010anisotropy} based on magnetic anisotropy data, the structural shape anisotropy of nanoclusters is manifested for the samples with metal concentration above the percolation threshold. Dependence of the shape anisotropy parameter $r_p/r_l$ can be reasonably well described by the linear function (Fig. \ref{Fig3}b).

\section{Conclusion}

In conclusion, in this non-destructive experiment we have verified that nanocomposites (Co$_{86}$Nb$_{12}$Ta$_2$)$_x$(SiO$_2$)$_{1-x}$ contain anisotropic metal clusters. For a whole concentration range from $x=0.193$ to $x=0.665$ in vicinity of the percolation threshold the nanoparticles are elongated along the growth direction perpendicular to the film plane. For the lowest concentration $x=0.193$ clusters with almost isotropic shape and average radius of 1 nm were observed, while for higher concentrations the average cluster length along the film normal is $1.5-7$ nm and the cross-section is $1-2$ nm. Therefore the grain shape anisotropy must be included into consideration of the magnetic properties of this system, such as ferromagnetic resonance studies \cite{lutsev2002spin,vyzulin2006special} or absolute values of the magnetic anisotropy constant evaluation \cite{stognei2010anisotropy}.

\begin{acknowledgements}
Authors thank European Synchrotron Radiation Facility for the provided experimental opportunities. We would like to acknowledge A. Vorobiev and O. Konovalov for the technical assistance and O.V. Stognei for provided samples.
\end{acknowledgements}

\bibliographystyle{spphys}       
\bibliography{jsnm.bib}   

\end{document}